# Correlations between Ground and Excited State Spectra of a Quantum Dot

D.R. Stewart[*†], D. Sprinzak[‡†], C.M. Marcus[†], C.I. Duruöz[*], J.S. Harris Jr.[*]

[*] Department of Electrical Engineering, Stanford University, Stanford CA 94305, USA
[†] Department of Physics, Stanford University, Stanford CA 94305, USA
[‡] Braun Center for Submicron Research, Department of Condensed Matter, Weizmann Institute of Science, Rehovot 76100, Israel

We have studied the ground and excited state spectra of a semiconductor quantum dot for successive numbers of electron occupancy using linear and nonlinear magnetoconductance measurements. We present the first observation of direct correlation between the $m^{th}$ excited state of the N electron system and the ground state of the N+m electron system for m up to 4. Results are consistent with a non-spin-degenerate single particle picture of the filling of levels. Electron-electron interaction effects are also observed as a perturbation to this model. Magnetoconductance fluctuations of ground states are shown as anticrossings where wavefunction characteristics are exchanged between adjacent levels.

Quantum dots (QDs) are small electrically conducting regions, typically a micrometer or less in size, containing from one to a few thousand electrons (1). Due to the small volume, the allowed electron energies within the dot are quantized, forming a discrete spectrum of quantum states, not unlike the energy levels of an atom. However, in contrast to the ordered shell structure of atomic spectra—a consequence of the spherically symmetric potential that binds the electrons—the *generic* energy spectrum of a QD (lacking any special symmetry) shows no shell structure, but instead possesses *universal* statistical properties associated with the underlying universality of quantum chaos (2).

Disordered or irregularly-shaped QDs are examples of so-called mesoscopic systems—small electronic structures intermediate in size between atoms and macroscopic (classical) objects that possess universal spectral and transport properties, independent of material, shape or disorder. In the past few years, a remarkable set of connections between mesoscopic systems, complex quantum systems such as heavy nuclei (where the statistical approach to spectra was first developed), quantum systems whose classical analogs are chaotic, and random matrix theory has emerged, providing deep theoretical insight into the generic properties of quantum systems (2). These connections are based principally on noninteracting single-particle spectral properties; only recently has attention been focused on mesoscopic systems in which interactions between particles and interference play equally important roles. It is in this context that the generic spectral features of multi-electron QDs are of great interest.

The electronic spectra of QDs are governed by the interplay of two energy scales: the Coloumb interaction or charging energy associated with adding a single electron to the dot, and the confinement energy associated with quantization due to the confining potential. In lateral semiconductor QDs the charging energy is typically an order of magnitude larger than the quantum confinement energy, leading one to expect that the quantum spectrum of the N+1 electron QD should be uncorrelated with the spectrum of the N electron QD (1).

Previous investigation of QD spectra via transport and capacitance spectroscopy has concentrated on analysis of the 'addition' energy spectrum composed of the ground states of the QD for successive numbers of electrons (1,3,4,5). Several experiments (6,7,8,9) have also probed the 'excitation' spectrum of quantum levels in the QD for *fixed* electron number using nonlinear conductance measurements. These measurements showed spectral features attributed to electron-electron interactions such as spin blockade (8) and clusters of resonances identified with a single excited state (9). Many-body calculations for few electron systems (N≤5) have explained some of the experimentally observed features in terms of spin and spatial selection rules (10), spectrally dominant center-of-mass excitation modes (11) and non-equilibrium effects (12). Very recently, ground and excited state spectra were both investigated in a circular, few-electron quantum dot (13).

We present an experimental study of correlations between ground state addition spectra and excitation spectra of an irregularly shaped QD containing several hundred electrons. We find that excitation and addition spectra for successive electron occupancies are remarkably correlated, agreeing in many respects with a non-interacting picture in which electrons simply fill the excited states, so that addition and excitation spectra coincide. Departures from this single particle model are also observed, giving insight into the electron-electron interaction strength. Additionally, we observe that spin degeneracy appears absent in the QD spectrum.

Measurements were performed on a quantum dot defined by applying ~ -0.3 V to Cr/Au electrostatic gates on the surface of a GaAs/AlGaAs heterostructure (inset Fig. 1B), depleting a two dimensional electron gas (2DEG) 900Å below the surface. The ungated 2DEG mobility and density were $1.4 \times 10^5$ cm$^2$/Vs and $2.0 \times 10^{11}$ cm$^{-2}$ at 4.2K. Differential conductance $g = dI/dV_{DS}$ of the QD in the Coulomb blockade (CB) regime (1) was measured in a dilution refrigerator using ac lock-in techniques with a 6 µV ac excitation added to a dc bias in the range ±1.5 mV. Measurements were made as a function of gate voltage ($V_G$), drain-source dc bias ($V_{DS}$) and magnetic field (B) applied perpendicular to the 2DEG plane. The experimental gate voltage was scaled to dot energy using both the nonlinear CB peak width (~ $eV_{DS}+3.5k_BT$) at finite $V_{DS}$ and independently by a fit of CB peak widths at $V_{DS} = 0$ as a function of temperature (14). The electron temperature in the dot was 90±10mK ($k_BT \sim 8$ µeV) as determined from the FWHM of linear CB peaks. The mean energy level spacing in the dot measured from the excited state spectra (discussed below) is $\Delta \sim 35$ µeV, providing an estimate of the dot area, $A = \pi \hbar^2 / m^* \Delta \sim 0.1$ µm$^2$ (m* is the electron effective mass). This area is consistent with the lithographic area allowing ~ 150 nm lateral depletion, and yields an occupancy of electrons in the dot N ~ 200. All magnetoconductance measurements were performed in the regime $g < 0.3$ e$^2$/h, and in the regime of single electron transport, i.e. $0 < |eV_{DS}| < E_C$, where $E_C \sim 730$ µeV is the classical charging energy of the QD measured from linear CB peak spacings.

A typical nonlinear differential conductance measurement through the QD as a function of $V_G$ and $V_{DS}$ is shown in Fig. 1A. At $V_{DS} = 0$ we observe the familiar CB peaks, approximately equally spaced in $V_G$. Increasing $V_{DS}$ results in broadening of the CB peaks to form multiple peak structures (6, 7) enclosing so-called 'Coulomb diamonds'. The central areas of the Coulomb diamonds (white in Fig. 1A) correspond to the blockade regime of zero conductance and fixed electron number. Dark stripes parallel to the Coulomb diamond edges in Fig. 1A are peaks in the differential conductance; each stripe represents the transmission resonance of a single QD level aligned with the source or drain Fermi levels (8). For positive $V_{DS}$ we identify the resonances parallel to the negative slope Coulomb diamond edge as unoccupied QD levels in resonance with the source, namely such peaks correspond to electrons tunneling into subsequent unoccupied states of the QD.

The first evidence of correlations between the excitation spectra of the N and N+1 electron systems can be seen in Fig. 1B, where the dark stripes of Fig. 1A are visible as multiple resonances on the left edge of each broadened CB peak. Each broadened CB peak shows a tall peak with one, two, three or four smaller peaks to its left. We identify the tall peak as the first, second, third or fourth excited state resonance of the N+3, N+2, N+1, or N electron system, respectively. At low temperature, $k_BT \ll \Delta$ ($\Delta/k_BT \sim 4$ in our dot) the peak height for each resonance is simply modeled as proportional to the overlap of the wave function in the dot with the source and drain wave functions (14).

Thus, the shift of the distinctive tall peak by one position in each successive excitation spectrum suggests that the particular electron wave function associated with this peak and the overall level structure of the dot near the Fermi surface are only weakly perturbed as electrons are removed one by one.

In order to confirm this correlation between excitation spectra of adjacent CB peaks, we follow the evolution of each resonance within a broadened CB peak as a function of magnetic field, B. The continuous evolution of resonance position and height with B yields a distinct signature or 'fingerprint' for each quantum level, and collectively for each excitation spectrum. Figure 2 shows three broadened CB peaks evolving with B, at fixed $V_{DS}$. We observe the striking effect that the fingerprints of the N-1, N, and N+1 excitation spectra display *shifted versions of the same level structure*; for each electron removed from the dot one extra level is visible at the top of each spectrum. This shifting agrees with a non-interacting description of the QD in which a fixed spectrum is filled one level at a time.

Figures 2B and 2C detail the labeling of the QD levels for the data of Figs. 1 and 2A. The dark stripes of Fig. 1A are identified by color as subsequent QD levels in resonance with the source fermi level. Figure 2B shows the state of the system after an electron has tunneled onto the QD from the source and before it has tunneled off to the drain, for the $V_G$ bias which aligns each ground state with the source. Increasing $V_G$ (lowering the dot potential well) brings higher excited states into resonance with the source. For the broadened CB peak separating the N-2 and N-1 Coulomb diamonds (red in Fig. 2A) we label these higher resonances as excited states of the N-1 electron system, and show the B field fingerprint of this N-1 excitation spectrum in Fig. 2A. This sequential labeling is understood to be approximately correct (we neglect non-equilibrium effects, which are not resolvable presumably due to thermal broadening (8, 12)). Negative differential conductance peaks are also observed in our data but will not be described here.

Further understanding of the QD spectral properties is obtained by comparing the fingerprints of ground state addition spectra measured at $V_{DS} = 0$ to that of neighboring excitation spectra, measured at finite $V_{DS}$. Six peaks, labeled N-3 to N+2 in Fig. 3A (where peak N represents the degeneracy between electron numbers N and N-1) show the fingerprints of several consecutive ground states, for different surface gate voltages and thus a different QD shape than in Figs. 1 and 2. Individual ground states show large peak height and position fluctuations, with the expected symmetry about B = 0. In Fig. 3C we collapse this group of levels, originally separated by $E_C$, by shifting each trace in energy (gate voltage) until they best align with adjacent levels. By doing so we assume a Coulomb interaction independent of B. Remarkably, such translation of the ground states produces a recognizable spectrum coherent over many levels, in which the fluctuations of height and position are visible as anti-crossings of neighboring levels. The specific signature of each anti-crossing is that two successive levels appear to trade both conductances and velocities ∂E/∂B as they pass through their point of closest approach (15). This *assembled* spectrum of fluctuating ground states appears to be composed of slowly varying wave functions *as followed through anti-crossings*, slightly perturbed into the measured anti-crossed level structure.

Finally, we compare the assembled ground state spectrum to measured excited state spectra of the same peaks. Figures 4A and 4B compare the $N^{th}$ CB peak at finite bias, showing structure that corresponds to the excited state spectrum of the N-electron QD, to the $N^{th}$ and $N+1^{th}$ peaks at zero-bias shifted in the same way as in the assembled spectrum of Fig. 3C. One can observe that the magnetoconductance fingerprint of the $N^{th}$ zero-bias CB peak matches the resonances at both the top and bottom of the $N^{th}$ finite bias CB peak, since all three are identified with the ground state of the N electron system. More significantly, the $N+1^{th}$ zero-bias CB peak (ground state) matches closely the *second* resonance (first excited state) of the $N^{th}$ finite bias peak in position, height and relative spacing between the levels. We emphasize that the observed correspondence between the $N+1^{th}$ ground state and the $N^{th}$ first excited state is trivially implied by a non-interacting electron model, but is not obvious in a strongly interacting system.

Despite the overall consistency of the observed addition and excitation spectra with a single-particle picture, there are some important departures that arise presumably due to electron-electron interactions. Figure 4C shows the finite bias structure corresponding to the excited states spectrum of the N-1 electron system, while Fig. 4D displays the $N-1^{th}$, $N^{th}$ and $N+1^{th}$ zero bias peaks shifted in energy from Fig. 3B to best match the N-1 excited state spectrum. The relative position of the $N^{th}$ and $N+1^{th}$ levels in Fig. 4D differs considerably from that in Fig. 4B. The $N+1^{th}$ ground state has been shifted from its alignment of Fig. 4B (same as Fig. 3C) until it is overlapping and even changing places with the $N^{th}$ ground state. This apparent 30 µeV shift of the $N+1^{th}$ level is comparable to the average level spacing of 35 µeV, and indicates that while levels may undergo an overall shift in energy as one electron is added, the level 'fingerprint' (position and height fluctuations in B) appears largely unchanged. Smaller, similar shifts in level spacings exist in almost all neighboring excitation spectra. Additionally, some resonances show a trend of broadening at higher excited state energies similar to (9).

Another departure from the simple single particle picture is the absence of spin-degenerate pairs of levels. Previous measurements on few-electron semiconductor QDs (5) and ultrasmall metal QDs (9) showed spin degenerate level spectra, whereas in-plane magnetic field measurements of a multi-electron semiconductor QD suggested no spin degeneracy (8). In our results the appearance of one new resonance in the excitation spectrum per electron removed from the QD indicates that energy levels in the dot are *not* spin degenerate. To estimate the energy splitting between spin paired levels, we examine the spectrum for levels with identical fingerprints. Figure 3B shows that each ground state level has a different fingerprint, implying none are spin paired. In Fig. 3C however, some of the slowly varying wave functions followed through anti-crossings do appear parallel suggesting that spin pairing may be visible in this underlying spectrum. We infer in either case that the energy splitting between spin paired wave-functions is larger than the mean single particle spacing, $\Delta \sim 35$ µeV. This energy splitting determines the scale of the spin-orbit (16) or electron-electron interaction responsible for the absence of degeneracy.

In conclusion, we have demonstrated for the first time that strong correlations exist between the QD energy level spectra of successive electron numbers in the dot, probed via magnetotransport measurements. The excitation spectra of adjacent CB peaks are found to be shifted versions of a very similar spectrum, with the addition of one excited state per electron removed from the dot. These results suggest a single particle model of the QD spectrum with no spin degeneracy. Departures from this single particle model are attributed to electron-electron interactions.

We thank B. Altshuler, D. Ralph and M. Heiblum for useful discussions. We thank S. Patel and A. Huibers for valuable help throughout the measurements. We gratefully acknowledge support from JSEP under Grant DAAH04-94-G-0058, the Army Research Office under Grant DAAH04-95-1-0331, the Office of Naval Research YIP program under Grant N00014-94-1-0622, and the NSF-NYI program. One of us (DS) acknowledges the support of MINERVA grant.

---

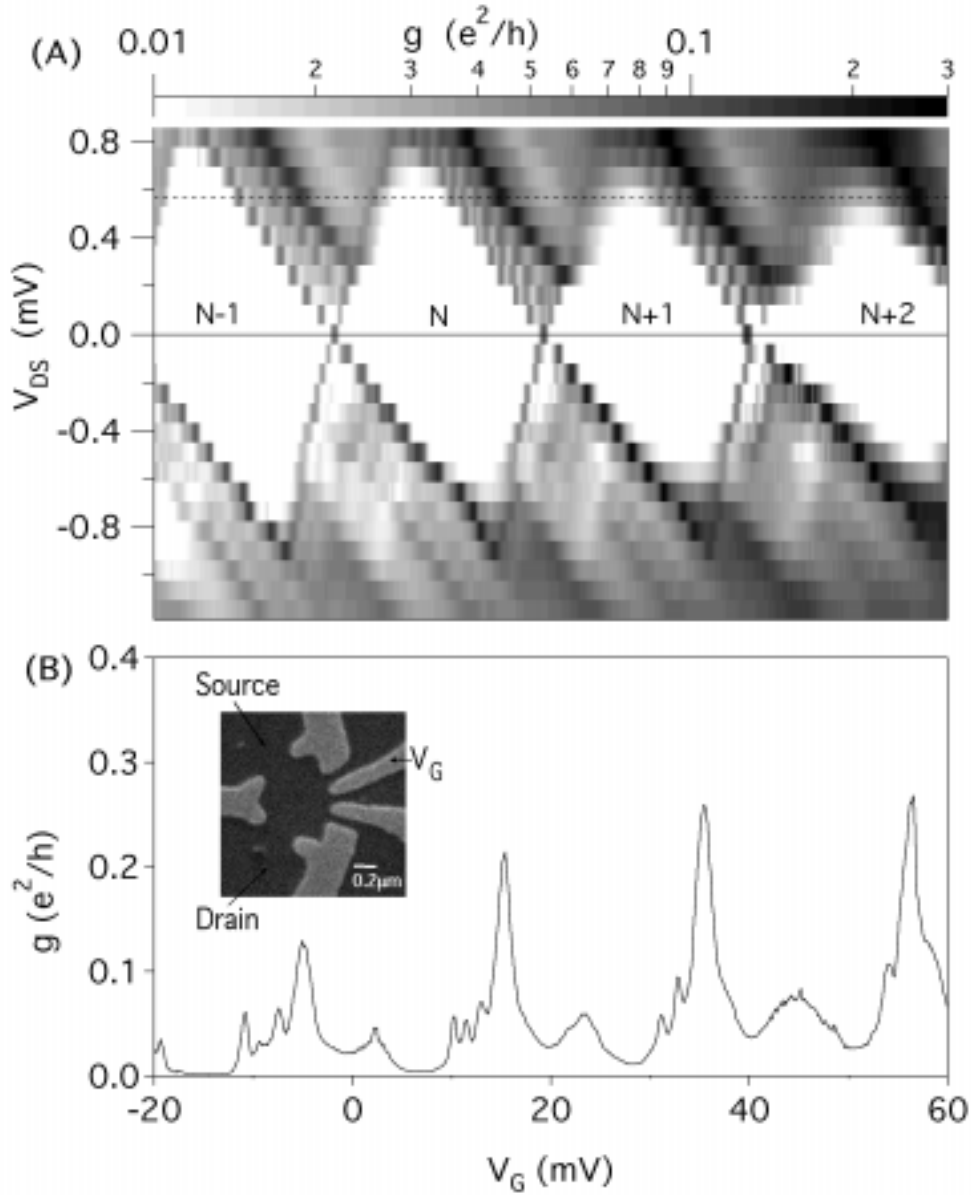

**Figure 1**. Coulomb diamonds: (A) Differential conductance, g, as a function of drain-source voltage, $V_{DS}$, and gate voltage, $V_G$, in a gray scale where black is large g. The white diamonds are blockade regions where electron number is fixed. Dark diagonal stripes (peaks) parallel to the diamond edges correspond to QD levels in resonance with the source or drain fermi levels. The magnetic field is 30 mT. (B) One trace from Fig. A at $V_{DS}$ = 570 µV. The distinctive high peak appears as the 2nd resonance from the left (1st excited state) on the rightmost CB peak and shifts to the 3rd, 4th and 5th (not resolvable) resonance (2nd, 3rd and 4th excited states) as the electron number decreases. Decreasing average g is attributed to capacitive coupling between $V_G$ and adjacent surface gates. Inset: A scanning electron micrograph of the QD studied.

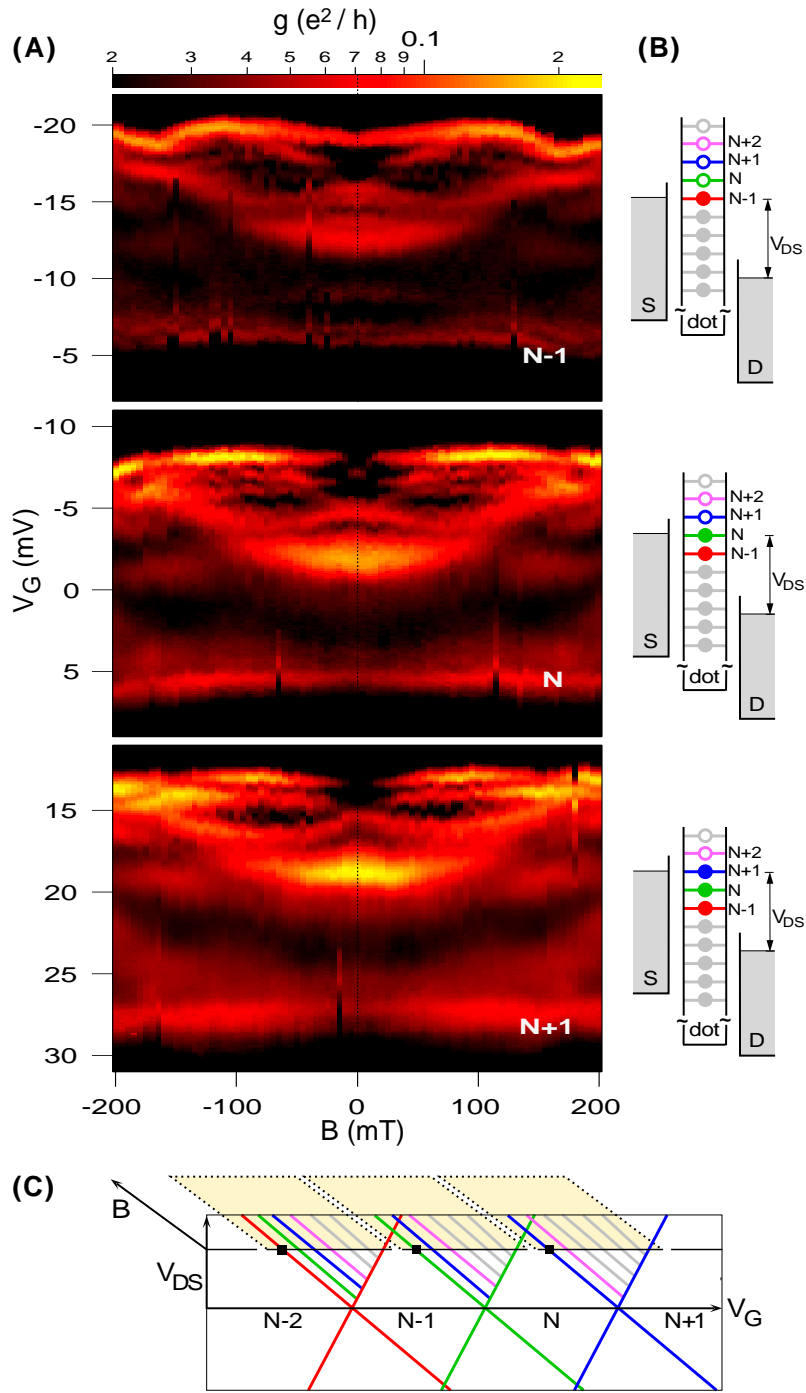

**Figure 2**. Excitation spectra correlations: (A) Differential conductance, g, of the three broadened CB peaks separating the N-2, N-1, N and N+1 Coulomb diamonds as a function of gate voltage, $V_G$, and applied magnetic field, B, in color scale where yellow is large g. Fixed drain-source bias $V_{DS}$ = 570 µV. The resonance pattern of each CB peak corresponds to the excitation spectrum of the N-1, N or N+1 electron QD. As each electron is removed a new resonance is introduced and the rest of the spectrum shifts by one level. Energy scale is inverted in this figure only. (B) Schematic energy diagrams of the source-QD-drain system, with N-1, N or N+1 electrons in the QD. Filled circles indicate occupied levels in the dot. The QD is illustrated with the N-1, N or N+1 level in resonance with the source fermi level. Increasing $V_G$ increases the dot potential well depth and brings higher levels into resonance with the source. (C) Coulomb diamond schematic, with colored stripes corresponding to colored levels in Fig. B, and black squares illustrating the specific gate bias $V_G$ of each diagram in Fig. B. The three yellow planes correspond to the data of Fig. A. Resonance stripes parallel to the positive slope Coulomb diamond edge are not illustrated as they are not clearly resolved in our data.

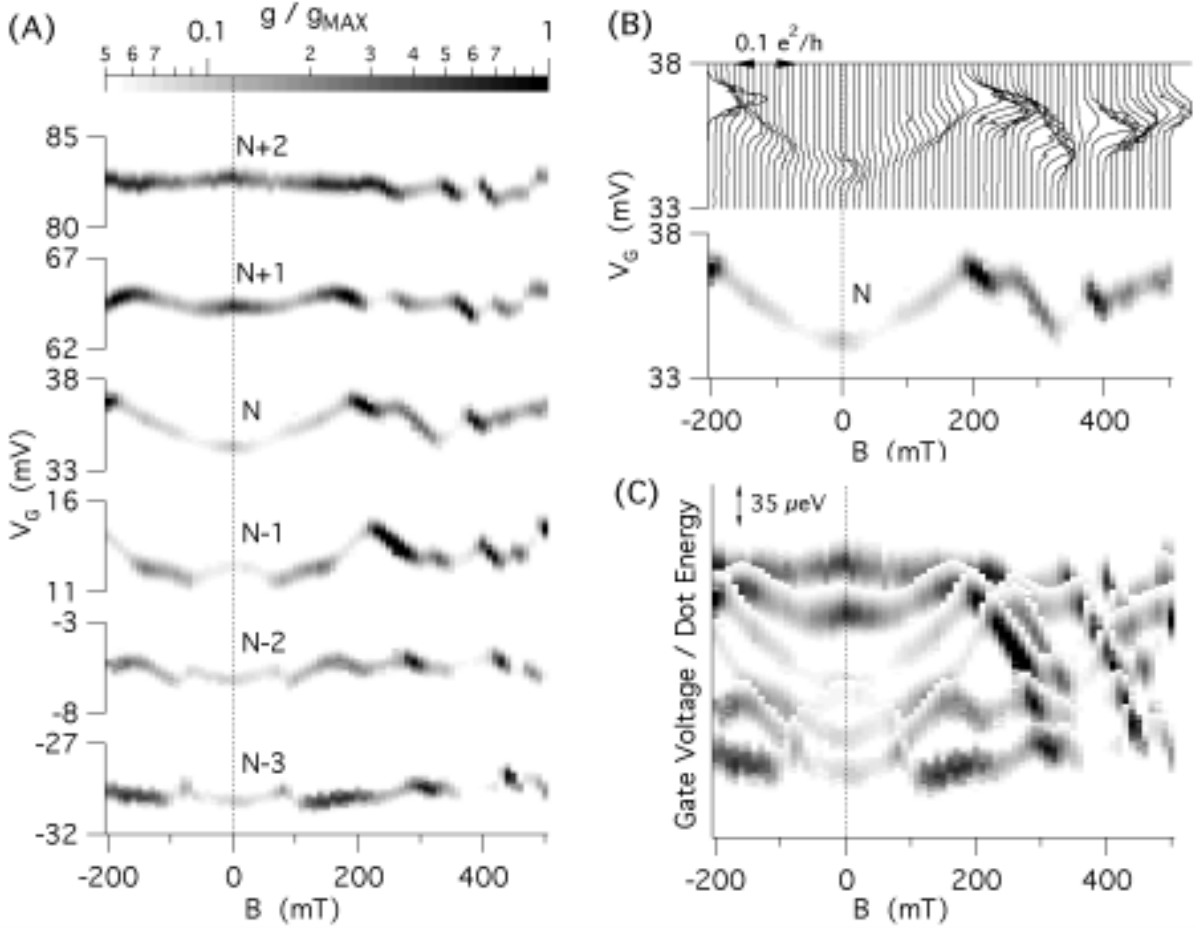

**Figure 3**. Ground state spectra: (A) Six adjacent CB peaks at $V_{DS} = 0$ in gray scale showing both position and height fluctuations as a function of B. The gray scale of g for each trace is normalized to the maximum g of that trace. (B) Trace N shown as raw data and the corresponding gray scale plot. (C) The six traces of Fig. A are shifted vertically to best align with each other, forming a spectrum with visible anti-crossings of adjacent levels.

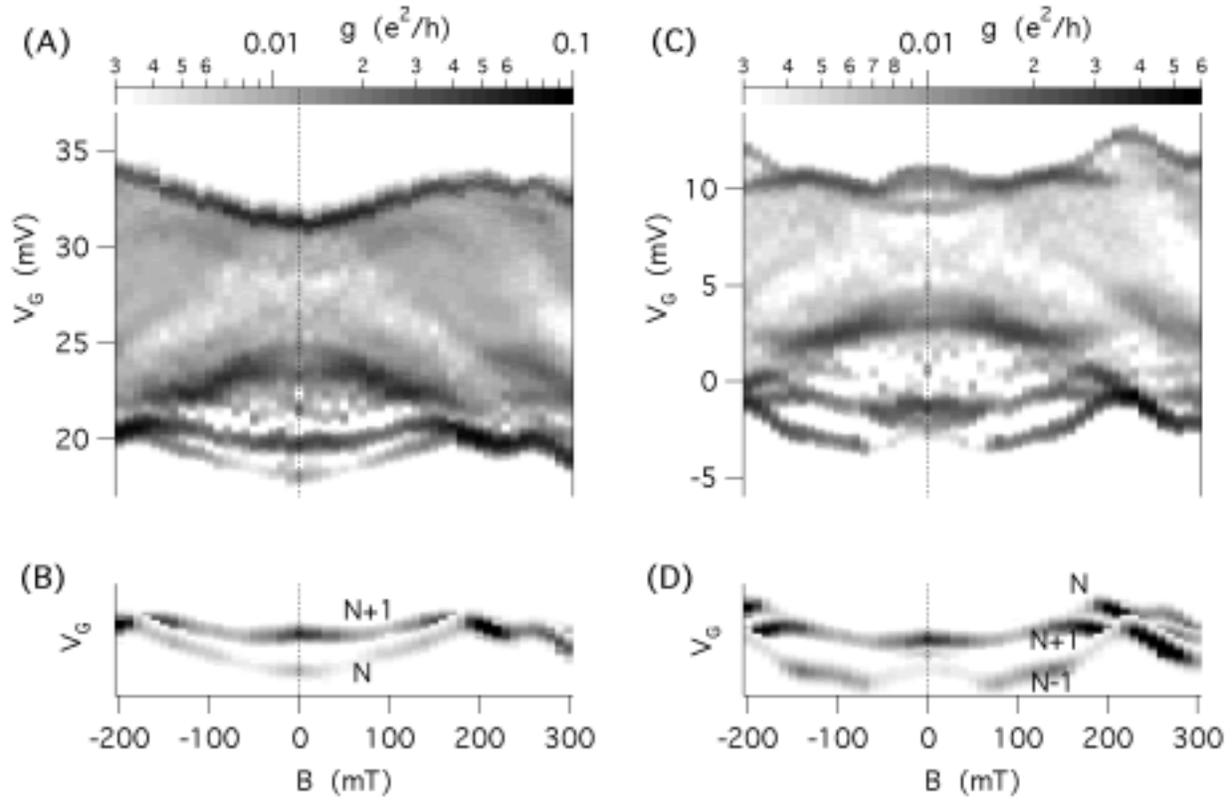

**Figure 4**. Comparing ground and excited state spectra: (A) The magnetic 'fingerprint' of the $N^{th}$ CB peak at $V_{DS}$ = 570 µV. Visible resonances at the bottom of the peak correspond to the excitation spectrum of the N electron QD. (B) The $N^{th}$ and $N+1^{th}$ zero bias CB peaks, shifted as in Fig. 3C. The $N^{th}$ peak matches both edges of the finite bias peak in A, and the $N+1^{th}$ peak matches the second resonance (first excited state) in Fig. A. A similar match between the excitation spectrum of the N-1 electron system and the N-1, N, N+1 ground states is done in (C) and (D). Note the relative position of the $N^{th}$ and $N+1^{th}$ peaks is different than in Fig. B.